\documentstyle[12pt]{article}
\begin{document}
\tolerance=5000
\def\pp{{\, \mid \hskip -1.5mm =}}
\def\cL{{\cal L}}
\def\be{\begin{equation}}
\def\ee{\end{equation}}
\def\bea{\begin{eqnarray}}
\def\eea{\end{eqnarray}}
\def\tr{{\rm tr}\, }
\def\nn{\nonumber \\}
\def\e{{\rm e}}
\def\D{{D \hskip -3mm /\,}}

\ \hfill
\begin{minipage}{3.5cm}
NDA-FP-60 \\
April 1999 \\
\end{minipage}

\vfill

\begin{center}
{\Large\bf Anomaly induced effective actions in even dimensions
and reliability of  $s$-wave approximation}

\vfill

{\sc Shin'ichi Nojiri}\footnote{
e-mail: nojiri@cc.nda.ac.jp}
and {\sc Sergei D. Odintsov$^{\spadesuit,\diamondsuit}$}\footnote{
e-mail: odintsov@mail.tomsknet.ru, odintsov@itp.uni-leipzig.de}

\vfill

{\sl Department of Mathematics and Physics \\
National Defence Academy, 
Hashirimizu Yokosuka 239, JAPAN}

\ 

{\sl $\spadesuit$ 
Tomsk State Pedagogical University, 634041 Tomsk, RUSSIA}

\ 

{\sl $\diamondsuit$ NTZ, University of Leipzig, Augustusplatz 10/11 \\
04109 Leipzig, GERMANY}

\

\vfill

{\bf abstract} 

\end{center}

The reduction of 4d matter-gravity theory to $S_2$, $H_2$ or 
$R_2$ leads to effective 2d dilatonic gravity with dilaton coupled 
matter. Spinors give the exceptional example of the theory which 
is conformally invariant in 4d as well as in 2d, after reduction. 
We find 4d and 2d conformal anomaly induced effective action (EA) 
for Majorana spinor. 
It is expected for some time that $s$-wave EA 
(i.e. the one for dilaton coupled 2d matter) 
is some ($s$-wave) approximation to 4d EA. We compare such 2d and 4d 
spinor EAs 
on the same gravitational background and argue that $s$-wave EA 
indeed qualitatively corresponds to no higher derivatives 
approximation for 4d EA.

\vfill


\newpage

It is well-known (see ref.\cite{H}) that spherical reduction of
gravity with matter leads to specific model of 2d dilaton gravity 
interacting with dilaton coupled matter. For example, 4d minimal 
scalar after reduction is described as 2d dilaton coupled scalar. 
The calculation of conformal anomaly as well as anomaly induced EA 
(which represents part of complete EA) for 2d dilaton coupled 
scalar has been done in refs.\cite{DilAnom}.

It is clear that such EA is given in the approximation where due to 
spherical symmetry of the problem one neglects the modes 
corresponding to spherical coordinates. That is why it is called 
EA in $s$-wave approximation. It is the hope that this EA in 
$s$-wave approximation may qualitatively 
describe 4d physics (BHs, wormholes or quantum cosmology). 
The attempts to apply it in the study of quantum properties of 
4d BHs may be found in refs.\cite{BH,BRM}.
In some sense such investigation is similar to description 
of quantum 2d BHs in CGHS and RST models 
of dilatonic gravity (for a very incomplete list of refs. see 
\cite{CGHS,RST,23}) but now 2d BH with dilaton is 
interpreted as 4d BH. 
Interchange of radial coordinate with time in above problem gives the 
way to apply EA in $s$-wave approximation in the study of quantum 
properties of 4d Kantowski-Sacks cosmologies \cite{KS} as it
was shown in refs.\cite{Gates}. There is clear analogy here with 
2d quantum cosmology as it follows from CGHS or RST-like models 
\cite{MR}. However, despite all these applications the precise 
correspondence 
between complete 4d EA (which is very difficult to find)
and EA in $s$-wave approximation (this EA is much easier to obtain)
is still lacking.

The purpose of this Letter will be to compare these two EAs and to 
clarify the reliability of $s$-wave approximation. Note that for 
conformally invariant 
matter one can easily get 4d anomaly induced EA on the arbitrary 
gravitational background. It is given with the accuracy up to 
conformally invariant functional which is numerical constant for 
some backgrounds.
On the same time the one-loop EA in $s$-wave approximation is 
known \cite{DilAnom} 
for minimal (non-conformal) scalar. Hence, scalar is not good example 
to make comparison of induced (complete) 2d and 4d EAs. (However, one can 
compare stress tensors from 4d minimal scalar and from corresponding 
EA in $s$-wave approximation).

We consider (Majorana) spinor. In this case 4d anomaly induced EA 
for conformal spinor is well-known. On the same time the 2d anomaly 
induced EA 
for dilaton coupled spinor which corresponds to $s$-wave reduction 
of conformal spinor has been found in ref.\cite{NNO}. Moreover, it actually 
gives complete EA (no conformally invariant functional appears)
as it was shown in ref.\cite{NNO}. We compare these EAs for the same
4d gravitational backgrounds and we show that in general they do 
not match even in the number of derivatives (higher derivatives 
4d EA versus $s$-wave EA which is of second order on derivatives). 
Nevertheless,we argue that $s$-wave EA may still be in qualitative 
agreement with complete 4d EA calculated in the approximation 
where higher derivatives terms may be neglected. Brief remarks on 
calculation of Hawking radiation from above actions are also given.

We will start from the Lagrangian of 4d dimensional (Majorana) 
spinors:
\be
\label{i}
L=\sum_{i=1}^N \bar\psi_i\gamma^\mu\nabla_\mu\psi^i\ .
\ee
It is well-known that such theory is conformally invariant. On the 
quantum level, the conformal invariance is broken, what leads to 
conformal anomaly \cite{M}
\be
\label{ii}
T=b\left(F + {2 \over 3}\Box R\right) + b'G
\ee
where square of the Weyl tensor is given by
$F=R_{\mu\nu\rho\sigma}R^{\mu\nu\rho\sigma} 
-2 R_{\mu\nu}R^{\mu\nu} + {1 \over 3}R^2 $
and Gauss-Bonnet invariant is 
$G=R_{\mu\nu\rho\sigma}R^{\mu\nu\rho\sigma} 
-4 R_{\mu\nu}R^{\mu\nu} + R^2$. 
For Majorana spinors
$b={3N \over 120(4\pi)^2}$, $b'=-{11 N \over 720 (4\pi)^2}$.

Having the explicit expression for conformal anomaly one can turn now 
to the calculation of the anomaly induced EA. Starting 
from the background with the metric
\be
\label{v}
g_{\mu\nu}=\e^{2\sigma}\bar g_{\mu\nu}
\ee
where $\bar g_{\mu\nu}$ is a fixed fiducial metric (reference) one 
can suppose that all dynamics is included to the conformal factor 
$\sigma$ dependence. Then, using the definition of stress-energy trace via 
some (gravitational) action with explicit help of Eq.(\ref{v}), 
one gets
\be
\label{vi}
T={1 \over \sqrt{-g}}{\delta W \over \delta\sigma}\ .
\ee
Substituting the metric (\ref{v}) to 
Eq.(\ref{ii}) and integrating over $\sigma$ one finds 
the anomaly induced EA \cite{R}.

This EA may be represented in the 
non-covariant local form \cite{R}
\bea
\label{vii}
W&=&\int d^4x \sqrt{-\bar g} \Bigl\{b \bar F \sigma 
+ 2b' \sigma\left[ \bar{\Box}^2 
+ 2 \bar R^{\mu\nu}\bar\nabla_\mu\bar\nabla_\nu \right. \nn
&& \left. - {2 \over 3}\bar R\bar{\Box} 
+ {1 \over 3}(\bar\nabla^\mu\bar R)\bar\nabla_\mu
\right]\sigma 
+ b'\sigma\left(\bar G -{2 \over 3}\bar\Box\bar R\right) \nn
&& -{1 \over 18}(b + b')\left[\bar R - 6 \bar{\Box} \sigma
- 6(\bar\nabla_\mu \sigma)(\bar\nabla^\mu \sigma) 
\right]^2\Bigr\}
\eea
where $\sigma$-independent terms (generalized integration constant) 
are not written explicitly. Hence, $W$ is not complete, the 
only $\sigma$-dependence is exactly taken into account. Note that 
one has to extract $\bar R^2$ term from the last term in (\ref{vii}) 
if to use the usual condition that $W$ vanishes at $\sigma=0$. 

It may be also presented in covariant but non-local form \cite{R}:
\bea
\label{viii}
W&=&-{1 \over 4b'}\int d^4x \sqrt{-g(x)}\int d^4y \sqrt{-g(y)}
\left[b F + b'\left(G - {2 \over 3}\Box R\right) \right]_x \nn
&& \times \left[ 2 \Box^2 + 4 R^{\mu\nu} \nabla_\mu\nabla_\nu 
- {4 \over 3} R \Box + {2 \over 3}\left(\nabla^\mu R\right)
\nabla_\mu\right]^{-1}_{xy} \nn
&& \times \left[b F + b'\left(G - {2 \over 3}\Box R\right) 
\right]_{y} - {1 \over 18}(b+b')\int d^4x \sqrt{-g} R^2 \ .
\eea
The total one-loop effective action is given as follows
\be
\label{ix}
\Gamma=W + \Gamma(\bar g_{\mu\nu})
\ee
where the second term (conformally invariant functional) cannot be 
found from the only conformal anomaly. Of course, 
second term (generalized integration constant) of $\Gamma$ is 
covariant one. It is only to show explicitly its conformal 
invariance we write the argument as tilded metric.

Let us give few explicit examples. 

\noindent
1. First of all, we consider conformally flat space, i.e., 
$\bar g_{\mu\nu}=\eta_{\mu\nu}$. In this exceptional case, the 
second term in Eq.(\ref{ix}) is the numerical constant which may be 
dropped away. Hence, in such case
\be
\label{x}
\Gamma=W=\int d^4x \Bigl\{2b' \sigma \bar{\Box}^2 \sigma 
- 2(b + b')\left( \bar{\Box} \sigma
 +(\bar\nabla_\mu \sigma)(\bar\nabla^\mu \sigma) 
\right)^2\Bigr\}\ .
\ee
To even simplify the problem, we suppose that $\sigma$ depends only 
on conformal time $\eta$. Then
\be
\label{xi}
W=V_3\int d\eta\left\{2b' \sigma \sigma'''' 
- 2(b + b')\left( \sigma'' + {\sigma'}^2 \right)^2\right\}\ .
\ee
Here $V_3$ is the (infinite) volume of 3-dimensional flat space, 
$'\equiv{d \over d\eta}$ and $\sigma=\ln\alpha$ where 
$\alpha(\eta)$ is the scale factor. It is known that semiclassical 
Einstein gravity (with quantum corrections described by the action 
(\ref{xi})) leads to the possibility of inflation \cite{SMM} even 
in the presence of dilaton \cite{BO}.

\noindent
2. Let us consider the situation discussed in ref.\cite{NO} 
when $\bar g_{\mu\nu}=\left(g_{\mu\nu}^{(2)}, 
g_{\alpha\beta}^\Omega\right)$ where $g_{\mu\nu}^{(2)}$ is an 
arbitrary two-dimensional metric and $g_{\alpha\beta}^\Omega$ is 
two-dimensional metric for the space with constant or zero curvature 
($S_2$, $H_2$ or $R_2$). Then supposing 
that $\sigma$ depends on coordinates from $g_{\mu\nu}^{(2)}$ only 
one gets
\bea
\label{12}
W&=&V_\Omega \int d^2x\sqrt{-g}\Bigl\{{b \over 3}\left[
\left(R^{(2)} + R_\Omega\right)^2 + {2 \over 3}R_\Omega R^{(2)} 
+ {1 \over 3}R_{\Omega}^2 \right] \sigma \nn
&& + b' \left[\sigma \left(
2\Box^2 + 4R^{(2)\mu\nu}\nabla_\mu \nabla_\nu 
-{4 \over 3}(R^{(2)} + R_\Omega)\Box \right.\right. \nn
&& \left.\left.
+ {2 \over 3}(\nabla^\mu R^{(2)})\nabla_\mu\right) \sigma\right]  
+ b'\left( 2R_\Omega R^{(2)} - {2 \over 3}\Box R^{(2)}
\right)\sigma \nn
&& -{1 \over 18}(b+b')\left( R^{(2)} + R_\Omega - 6 \Box\sigma 
- 6\nabla^\mu\sigma\nabla_\mu\sigma \right)^2 \Bigr\} 
\eea
$W$ takes the simplest form when 
$g_{\alpha\beta}^\Omega=\eta_{\alpha\beta}$, i.e. $R_\Omega=0$. 
However, even in this case there is an additional, conformally 
invariant functional $\Gamma$ which is unknown and which should 
be added to $W$ to form  the complete one-loop effective action 
(\ref{ix}). This unknown functional may be found as an expansion 
(see \cite{NO}) but in this case conformal invariance may be lost. 
Hence, we presented anomaly induced EA in its 
explicit form for two metrics which come naturally from 
cosmological and black hole considerations.

At the next point, we consider the same Majorana spinor 
Lagrangian (\ref{i}). Assume the 4d compactified spacetime
\be
\label{xiii}
ds^2 = g_{\mu\nu}dx^\mu dx^\nu + r_0^2 \e^{-2\phi}d\Omega_{2d}
\ee
where $d\Omega_{2d}$ corresponds to 2d unit sphere $S_2$: 
$d\Omega_{2d}=d\theta^2 + \sin^2\theta d\varphi^2$ 
or 2d flat space $R_2$: $d\Omega_{2d}=dydz$ 
or unit hyperboloid $H_2$: 
$d\Omega_{2d}=d\theta^2 + \sinh^2\theta d\varphi^2$
and $\phi$ depends on coordinates corresponding to 
$g_{\mu\nu}$. 
Then we find 
$\sqrt{-g^{(4)}}=r_0^2 \e^{-2\phi}\sin\theta\sqrt{-g^{(2)}}$ 
for $S_2$, $\sqrt{-g^{(4)}}=r_0^2 \e^{-2\phi}\sqrt{-g^{(2)}}$ 
for $R_2$ and 
$\sqrt{-g^{(4)}}=r_0^2 \e^{-2\phi}\sinh\theta\sqrt{-g^{(2)}}$
for $H_2$. The integration of $\varphi$ 
($0\leq\varphi<2\pi$) and $\theta$ ($0\leq\theta\leq\pi$ for 
sphere and $0\leq\theta<\infty$ for hyperboloid) or $(y,z)$ for 
$R_2$ give $4\pi$ for $S_2$ and an infinite constant 
for $R_2$ or $H_2$.

In the metric ansatz (\ref{xiii}), not all the spin connections 
vanish. The non-vanishing spin connections do not appear in 
the action of the Majorana fermion $\psi$. 
Therefore we find in the metric (\ref{xiii}) 
$\bar\psi\gamma^m D_m\psi = \bar\psi\gamma^\mu\nabla_\mu\psi$. 
Here $m=0,1,2,3$ ($\mu=0,1$) and $D_m$ is the covariant derivative
in 4d. We should note, however, that $N$ Majorana fermions in 4d 
corresponds to $2N$ Majorana fermions in 2d. 

One finds
\be
\label{avi}
\int d^4x \sqrt{-g^{(4)}}\sum_{i=1}^N\bar\psi_i\gamma^m D_m\psi_i 
= Cr_0^2 \int d^2 x \sqrt{-g^{(2)}}\e^{-2\phi}\sum_{i=1}^{2N}
\bar\psi_i\gamma^\mu\nabla_\mu\psi_i\ .
\ee
Here $C=4\pi$ for $S_2$ and $C$ is an infinite 
constant for $R_2$ or $H_2$. Then if redefine $\psi_i$ by
$\psi_i = {\psi^{(2)}_i \over r_0}$, we obtain
\be
\label{xiv}
\int d^4x \sqrt{-g^{(4)}}\sum_{i=1}^N\bar\psi_i\gamma^m D_m\psi_i 
= C \int d^2 x \sqrt{-g^{(2)}}\e^{-2\phi}\sum_{i=1}^{2N}
\bar\psi^{(2)}_i\gamma^\mu\nabla_\mu\psi^{(2)}_i\ .
\ee
Here $\psi^{(2)}$ is the usual spinor 
(with mass dimension 1/2) in 2d. And we abbreviate 
the suffix ``(2)'' in the following if there is no confusion.
The action 
(\ref{xiv}) describes 2d dilaton coupled spinor which is still 
conformally invariant one. The corresponding 2d dilaton dependent 
conformal anomaly has been calculated in ref.\cite{NNO} as follows
\be
\label{xv}
T=c\left[{1 \over 2}R + 2\triangle \phi \right]\ .
\ee
where $c={N \over 12\pi}$.  Notice that this is an anomaly 
of two-dimensional quantum spinor.
The anomaly induced EA in this case has been also found in 
ref.\cite{NNO} in non-local, covariant form. It is more convienient for 
our purposes to use local, non-covariant form of anomaly induced 
EA. 

We suppose that two-dimensional metric in Eq.(\ref{xiv}) is chosen 
in the same way as 4d metric in (\ref{v}), i.e. 
$g_{\mu\nu}=\e^{2\sigma}\bar g_{\mu\nu}$. Then
$R=\e^{-2\sigma}\left(\bar R - \bar\triangle\sigma\right)$, 
$\triangle \phi = \e^{-2\sigma}\bar\triangle\phi$. 
Then Eq.(\ref{vi}) looks as 
\be
\label{xvii}
{\delta W \over \delta \sigma}= c \left\{{1 \over 2}
\left(\bar R - 2 \bar\triangle\sigma \right) + 2 \bar\triangle 
\phi\right\}\ .
\ee
Integrating over $\sigma$, one finds the anomaly induced EA 
in generalized $s$-wave approximation (we call it generalized 
$s$-wave one because 
 4d metric is compactified to $S_2$ ,or to $H_2$, or to
$R_2$)
\be
\label{xviii}
W=c\int d^2x \sqrt{-\bar g}
\left\{{1 \over 2}\bar R\sigma - {1 \over 2}\sigma
\bar\triangle\sigma + 2\sigma\bar\triangle\phi\right\}\ .
\ee
The corresponding covariant, non-local expression looks as 
\cite{NNO}
\be
\label{xix}
W=-{c \over 4}
\int d^2x \sqrt{-g}\left\{{1 \over 2}R{1 \over \triangle}R
+ 4\phi R\right\}\ .
\ee
The remarkable property of dilaton coupled spinor is that $W$ 
gives the complete one-loop EA\cite{NNO} (up to 
non-essential constant) on the arbitrary dilaton-gravitational 
background. No conformally invariant functional is necessary 
for 2d dilaton coupled spinor unlike 4d case.

Notice also\cite{NNO} that the effective action (\ref{xix}) for 
2d dilaton coupled spinor gives natural realization of RST model 
\cite{RST} which represents the extension of CGHS model \cite{CGHS} 
on quantum level.  

Some remarks are in order. We got the effective action 
(\ref{xviii}), (\ref{xix}) in generalized $s$-wave approximation. 
It means that we first did reduction (\ref{xiii}) of classical 
matter action (\ref{xiv}). Then we neglected the dependence of 
quantum EA from two coordinates ($d\Omega$). The 
``memory'' on 4d space came through dilatonic factor which 
appeared in Eq.(\ref{xiv}). The Lagrangian (\ref{xiv}) describes 
now 2d quantum theory.  $s$-wave EA comes from 
this 2d quantum theory. As reduction and quantization 
do not commute, generally speaking,
it is clearly that such $s$-wave EA is not the same as 
4d EA. Nevertheless, it is usual hope that such 
$s$-wave effective action still qualitatively well
(in some sense) describes 4d 
physics. Moreover, it should have some relation with true effective 
action (for an introduction, see \cite{BOS}) which is not available 
usually. For example, the standard guess is that $s$-wave EA gives 
some approximation for complete 4d EA. 

Our purpose now will be to calculate  EA (\ref{xviii}), 
(\ref{xix}), in $s$-wave approximation on the same two backgrounds 
where 4d anomaly induced EA(\ref{vii}), (\ref{viii}) 
is calculated and to compare them.

First of all, we consider conformally flat space, i.e., 2d metric 
$\bar g_{\mu\nu}=\eta_{\mu\nu}$, $\sigma=\sigma(\eta)$ and in
Eq.(\ref{xiii}) $d\Omega$ corresponds to 2d flat space while 
$\phi=-\sigma$. Anomaly induced action (\ref{xviii}) at such 
conditions is
\be
\label{xx}
W=cV_1\int d\eta \left\{-{5 \over 2}\sigma\sigma''\right\}\ .
\ee
As one can see there is no even qualitative correspondence with 4d
anomaly induced EA (\ref{xi}) where we have only the derivatives of 
fourth order while in (\ref{xx}) of second order. However, it is not 
the full story. That is due to our background choice one gets no
second order derivatives in 4d anomaly induced EA.

It is not difficult to find anomaly induced 2d EA for the second 
example background discussed in 4d case above. Then
\be
\label{aivaa}
W=c\int d^2x \sqrt{-g}
\left\{{1 \over 2} R^{(2)}\sigma - {5 \over 2}\sigma
\triangle\sigma \right\}\ .
\ee
This again gives complete one-loop EA. Again,
this anomaly induced EA and 4d EA (\ref{12}) 
do not agree even in the number of derivatives. 
The explicit correspondence 
is difficult to check as in 4d case one has unknown conformally 
invariant functional 
completing EA. However, imagine that we are in a situation where 
one can argue that higher derivatives terms in 4d EA may be omitted 
(for example, adiabatic expansion, i.e. slowly varying fields). 
Then in such approximation 
one can see that only terms with two derivatives survive in 
Eq.(\ref{12}). Moreover, the structure of these terms remarkably 
repeats the structure of above 2d EA. Both these EAs are proportional
to number of spinors. It is clearly seen from Eq.(\ref{12}) 
(we take $R_{\Omega}=2$)
\bea
\label{12b}
W&\sim&V_\Omega \int d^2x\sqrt{-g}\Bigl\{{16b \over 9}
\left(R^{(2)} + 1\right) \sigma 
+ b' \left(-{8 \over 3}\sigma \triangle\sigma 
+ 4 R^{(2)} \sigma\right) \nn
&& -{2 \over 9}(b+b')\left( R^{(2)} - 6 \triangle\sigma 
- 6\nabla^\mu\sigma\nabla_\mu\sigma \right) \Bigr\} 
\eea
where higher derivative and constant terms are dropped.
Trivial dimensional transformation should be done in order to
present it in 2d form. 
As we see signs and coefficients do not match, of course. 
One can expect that 
after the calculation of the conformally invariant functional 
in complete 4d EA the approximation may become better.

Let us briefly comment on Hawking radiation.asit follows from above
actions.
We now work in the conformal gauge, where the metric has the 
following form:
\be
\label{HR01}
ds^2=\e^{2\rho}\left(-dt^2
+ ds^2\right) + r_0^2\e^{-2\phi} d\Omega^2 \ .
\ee
The Schwarzschild metric 
\be
\label{HR1}
ds^2=-\left(1-{r_0 \over r}\right)dt^2
+\left(1-{r_0 \over r}\right)^{-1}dr^2 + r^2 d\Omega^2 \ .
\ee
can be transformed into the metric in the conformal gauge 
(\ref{HR01}) by changing the coordinate 
\be
\label{HR2}
s=r+r_0\ln\left({r \over r_0}-1\right)\ .
\ee
Then the metric in (\ref{HR1}) has the following form:
\be
\label{HR3}
\e^{2\rho}=\left(1-{r_0 \over r(s)}\right)\ ,\quad
\e^{-2\phi}=r(s)^2 \ .
\ee
 Substituting (\ref{HR3}) into the quantum part of 
the stress-energy tensor $T^{\mu\nu}$ given by the effective 
action (\ref{xix}) induced from 2d anomaly, we obtain
\bea
\label{HR03}
T_{+-}&=&{c \over 16}\left(1 - {r_0 \over r}\right)
\left({2 \over r^2} - {5r_0\over r^3}\right) \nn
T_{\pm\pm}&=&{c \over 16}
\left({3r_0^2 \over 4r^4} - {2r_0 \over r^3} + {1 \over r^2}
\right) + f^\pm (x^\pm) \ .
\eea
Here $x^\pm = t\pm s$ and $M$. 
The functions $f^\pm (x^\pm)$ appear due to the non-locality 
in (\ref{xix}) and should be fixed by the boundary condition. 
Here we assume $T_{\pm +}=0$ at the past null infinity 
($x^-\rightarrow -\infty$) and $T_{\pm -}=0$ at the past horizon 
($x^+\rightarrow -\infty$). 
Then we find 
\be
\label{HR04}
f^+(x^+)=0\ ,\quad f^-(x^-)={c \over 64r_0^2}\ .
\ee 
Now one can estimate Hawking radiation (which 
is proportional to particles number) at the future null infinity 
$x^+\rightarrow +\infty$):
\be
\label{HR05}
T_{--}\rightarrow {c \over 64r_0^2}\ .
\ee
This result looks simple enough in 2d language.

The effective action (\ref{viii}) induced from 4d anomaly is also 
non-local and there is a possibility to generate the Hawking 
radiation. 
In the Schwarzschild metric, the curvatures have 
the following form:
\be
\label{HR4}
R=R_{\mu\nu}=0\ ,\quad R_{\mu\nu\rho\sigma}R^{\mu\nu\rho\sigma}
={12 r_0^2 \over r^6}\ ,
\ee
Then we find an expression of the trace anomaly (\ref{ii}):
\be
\label{HR5}
T={12\left(b+b'\right)r_0^2 \over r^6}\ .
\ee
The arguments on calculation of Hawking radiation using 
4d conformal anomaly in the 
Schwarzschild metric and the conservation law 
${T^{\mu\nu}}_{;\nu}=0$ have already been given in \cite{CF}. 
The results should be equivalent to that based on the complete 
non-local effective action (\ref{viii}). 
It was shown in \cite{CF} that there remain two integration 
constants and an unknown function 
which is relevant to the Hawking radiation but cannot be fixed 
from the conservation law and the trace anomaly. They could be 
fixed defining the vacuum state and/or the Hawking radiation 
itself.

Hence our main conclusion is that $s$-wave EA may be still relevant 
to 4d physics. However, it should be possible only for gravitational 
backgrounds where no higher derivatives terms approximation to EA is 
valid. 

\ 

\noindent
{\bf Acknowledgments}. SDO would like to thank S.W. Hawking 
and P. van Nieuwenhuizen for discussion of related questions. 
The research by
SDO was partially supported by a RFBR Grant N\,99-02-16617,
by Saxonian Min. of Science and Arts and by Graduate College
``Quantum Field Theory" at Leipzig University.


\begin{thebibliography}{99}
\bibitem{H}  P. Hajicek, {\sl Phys.Rev.} {\bf D30} (1984) 1178;
P. Thomi, B. Isaak and P. Hajicek, {\sl Phys.Rev.} {\bf D30} 
(1984) 1168.
\bibitem{DilAnom} E. Elizalde, S. Naftulin 
and S.D. Odintsov,  {\it Phys.Rev.} {\bf D49} (1994) 2852, 
hep-th/9308020;  
T. Chiba and M. Siino, {\it Mod.Phys.Lett.} {\bf A12} (1997) 709; 
R. Bousso and S.W. Hawking, {\it Phys.Rev.} 
{\bf D56} (1997) 7788, hep-th/9705236;
S. Nojiri and S.D. Odintsov, {\it Mod.Phys.Lett.} 
{\bf A12} (1997) 2083, hep-th/9706009 and {\it Phys.Rev}
 {\bf D57} (1998) 2363, hep-th/9706143;
Shoichi Ichinose, {\it Phys.Rev.} {\bf D57} (1998) 6224, 
hep-th/9707025;
A. Mikovic and V. Radovanovic, {\it Class.Quant.Grav.} 
 {\bf 15} (1988) 827,hep-th/9706066;
W. Kummer, H. Liebl and D.V. Vassilevich, 
{\it Mod.Phys.Lett.} {\bf A12} (1997) 2683, hep-th/9707041;
J.S. Dowker, {\it Class.Quant.Grav.} {\bf 15} (1998) 1881, 
hep-th/9802029 ;
S. Ichinose and S.D. Odintsov, {\it Nucl.Phys.} {\bf B539} 
(1999) 643, hep-th/9802043.
\bibitem{BH} R. Bousso and S.W. Hawking, {\it Phys.Rev.} 
{\bf D57} (1998) 2436, hep-th/9709224; 
S. Nojiri and S.D. Odintsov, {\it Phys.Rev.} {\bf D59} (1999) 044003,
hep-th/9806055;
R. Bousso, {\it Phys.Rev.} {\bf D58} (1998) 083511, hep-th/9805081.
\bibitem{BRM} M. Buric, V. Radovanovic and A. Mikovic, 
{\it Phys.Rev.} {\bf D59} (1999) 084002, gr-qc/9804083;
F.C. Lombardo, F.D. Mazzitelli and J.G. Russo,
{\it Phys.Rev.} {\bf D59} (1999) 064007, gr-qc/9808048;
R. Balbinot and A. Fabbri, {\it Phys.Rev.} {\bf D59} (1999) 044031,
hep-th/9807123 and gr-qc/9904034; 
W. Kummer and D.V. Vassilevich, hep-th/9811092.
\bibitem{23}  
 S.P. de Alwis, {\it Phys.Lett.} {\bf B289} (1992) 278;
A. Bilal and C. Callan, {\it Nucl.Phys.} {\bf B394} (1993) 73;
S. Nojiri and I. Oda, {\it Phys.Lett.} {\bf B294} (1992) 317;
{\it Nucl.Phys.} {\bf B406} (1993) 499;
T. Banks, A. Dabholkar, M. Douglas and M. O`Loughlin, 
{\it Phys.Rev.} {\bf D45} (1992) 3607;
R.B. Mann, {\it Phys.Rev.} {\bf D47} (1993) 4438;
D. Louis-Martinez and G. Kunstatter, {\it Phys.Rev.} {\bf D49} 
(1994) 5227;
J. Polchinski and A. Strominger, {\it Phys.Rev.} {\bf D50} (1994) 7403;
 T.Klobsch and T.Strobl, {\it Class.Quant.Grav.} {\bf 13} (1996) 965;
 S. Bose, L. Parker and Y. Peleg, 
{\it Phys.Rev.} {\bf D52} (1995) 3512; for a review,see
A. Strominger, Les Houches lectures on black holes, 
hep-th/9501071.
\bibitem{KS} R.Kantowskii and R. Sacks, J.Math.Phys.7(1967)2315
\bibitem{Gates} T. Kadoyoshi, S. Nojiri and S.D. Odintsov, 
{\it Phys.Lett.} {\bf B425} (1998) 255, hep-th/9712015; 
S.J. Gates, T. Kadoyoshi, S. Nojiri and S.D. Odintsov,
{\it Phys.Rev.} {\bf D58} (1998) 084026, hep-th/9802139;
S. Nojiri, O. Obregon, S.D. Odintsov and K.E. Osetrin,
to appear in {\it Phys.Rev.} {\bf D}, hep-th/9902035
 and {\it Phys.Lett.} {\bf B449} (1999) 173.
\bibitem{MR} F. Mazzitelli and J. Russo, 
{\it Phys.Rev.} {\bf D47} (1993) 4490;
A. Fabbri and J. Russo, {\it Phys.Rev.} {\bf D53} (1996) 6995;
W.T. Kim and M.S. Yoon, {\it Phys.Rev.} {\bf D58} (1998) 084014;
S. Bose and S. Kar, {\it Phys.Rev.} {\bf D56} (1997) 4444.
\bibitem{M} S. Deser, M.J. Duff and C.J. Isham, , {\it Nucl.Phys.} 
{\bf B111} (1976) 45, for a review, see M. Duff, 
{\it Class.Quant.Grav.} {\bf 11} (1994) 1387.
\bibitem{R} R.J. Riegert, {\it Phys.Lett.} {\bf B134} (1984)56;
E.S. Fradkin and A. Tseytlin, {\it Phys.Lett.} {\bf B134} (1984) 187;
I.L. Buchbinder, S.D. Odintsov and I.L. Shapiro, {\it Phys.Lett.}
{\bf B162} (1985) 92, 
I. Antoniadis and E. Mottola, {\it Phys.Rev.}  {\bf D45}
(1992) 2013; S.D. Odintsov, {\it Z.Phys.} {\bf C54} (1992) 531.
for a recent review see D. Anselmi, hep-th/9903059.
\bibitem{SMM} A. Starobinsky, {\it Phys.Lett.} {\bf B91} (1980) 99;
S.G. Mamaev and V.M. Mostepanenko, {\it JETP} {\bf 51} (1980) 9.
\bibitem{BO} I. Brevik and S.D. Odintsov, hep-th/9902184, 
to appear in {\it Phys.Lett.} {\bf B}.
\bibitem{NO} S. Nojiri and S.D. Odintsov, {\it Phys.Rev} {\bf D59} 
(1999) 044026, hep-th/9804033.
\bibitem{NNO} P. van Nieuwenhuizen, S. Nojiri and S.D. Odintsov, 
hep-th/9901119, to appear in {\it Phys.Rev} {\bf D}.
\bibitem{RST} J.G. Russo, L. Susskind and L. Thorlacius, 
{\sl Phys. Lett.} {\bf B292} (1992) 13.
\bibitem{CGHS} C.G. Callan, S.B. Giddings, J.A. Harvey 
and A. Strominger, {\it Phys. Rev.} {\bf D45} (1992) 1005.
\bibitem{BOS} I.L. Buchbinder, S.D. Odintsov and I.L. Shapiro,
{\sl Effective Action in Quantum Gravity}, IOP Publishing,
Bristol and Philadelphia, 1992.
\bibitem{CF} S.M. Christensen, S.A. Fulling, {\it Phys.Rev} 
{\bf D8} (1977) 2088.
\end{thebibliography}
\end{document}